\RequirePackage[2020-02-02]{latexrelease}
\documentclass[aps,prl,reprint,superscriptaddress,noeprint,showkeys]{revtex4-2}

\usepackage[hidelinks]{hyperref}
\usepackage{xcolor}
\hypersetup{
    colorlinks,
    linkcolor={blue},
    citecolor={blue},
    urlcolor={blue},
}
\usepackage{graphicx}
\usepackage{amsmath,braket,graphicx,mathtools,amssymb}
\usepackage{amsfonts}

\newcommand{\WTe}{$\mathrm{WTe_2}$}

\newcommand{\TaIrTe}{$\mathrm{TaIrTe_4}$}

\newcommand{\boldjr}{$\boldsymbol{J}(\boldsymbol{r})$}

\begin{document}


\title{Visualizing bulk and edge photocurrent flow in anisotropic Weyl semimetals}

\author{Yu-Xuan Wang}
\author{Xin-Yue Zhang}
\author{Chunhua Li}
\author{Xiaohan Yao}
\affiliation{Department of Physics, Boston College, Chestnut Hill, MA, 02467, USA}
\author{Ruihuan Duan}
\affiliation{CINTRA CNRS/NTU/THALES, UMI 3288, Nanyang Technological University, Singapore, Singapore.}
\author{Thomas K. M. Graham}
\affiliation{Department of Physics, Boston College, Chestnut Hill, MA, 02467, USA}
\author{Zheng Liu}
\affiliation{CINTRA CNRS/NTU/THALES, UMI 3288, Nanyang Technological University, Singapore, Singapore.}
\affiliation{School of Materials Science and Engineering, Nanyang Technological University, Singapore, Singapore.}
\affiliation{School of Electrical and Electronic Engineering \& The Photonics Institute, Nanyang Technological University, Singapore, Singapore.}
\author{Fazel Tafti}
\author{David Broido}
\author{Ying Ran}
\author{Brian B. Zhou}
\email{brian.zhou@bc.edu}
\affiliation{Department of Physics, Boston College, Chestnut Hill, MA, 02467, USA}

\date{\today}

\begin{abstract}
Materials that rectify light into current in their bulk are desired for optoelectronic applications. In inversion-breaking Weyl semimetals, bulk photocurrents may arise due to nonlinear optical processes that are enhanced near the Weyl nodes. However, the photoresponse of these materials is commonly studied by scanning photocurrent microscopy (SPCM), which convolves the effects of photocurrent generation and collection. Here, we directly image the photocurrent flow inside the type-II Weyl semimetals \WTe{} and \TaIrTe{} using high-sensitivity quantum magnetometry with nitrogen-vacancy center spins. We elucidate an unknown mechanism for bulk photocurrent generation termed the anisotropic photothermoelectric effect (APTE), where unequal thermopowers along different crystal axes drive intricate circulations of photocurrent around the photoexcitation. Using simultaneous SPCM and magnetic imaging at the sample's interior and edges, we visualize how the APTE stimulates the long-range photocurrent collected in our Weyl semimetal devices through the Shockley-Ramo theorem. Our results highlight an overlooked, but widely relevant source of current flow and inspire novel photodetectors using homogeneous materials with anisotropy.

\end{abstract}

\maketitle

For directional photocurrent flow absent of bias voltage, symmetry breaking is an essential ingredient  \cite{Cao2016,Ma2019,Sunku2020,Akamatsu2021,Jiang2021}. In device applications, symmetry is commonly broken by joining dissimilar materials or different dopings of the same material, driving photocurrent flow through a difference in their Seebeck coefficients \cite{Gabor2011,Buscema2013} or through the built-in electric field at a p-n junction \cite{Lee2014,Deng2014}. Generating photocurrents throughout a single homogeneous material can, however, be advantageous. One such intrinsic mechanism is the bulk photovoltaic effect (BPVE), a nonlinear optical process exhibited by non-centrosymmetric crystals \cite{Sipe2000,Tan2016,Morimoto2016,Ma2021}. The BPVE generates a steady photocurrent from the asymmetry in the electron wavefunctions before and after photoexcitation, allowing the polarization of light to control the photocurrent directionality.
 
Nonmagnetic Weyl semimetals, which feature topological band touchings preserved by the breaking of inversion symmetry, are compelling candidates to host the BPVE. Recent insight has cast the BPVE in terms of the quantum geometric properties of the band structure, predicting an enhanced response for low-energy excitation due to the diverging Berry curvature near the Weyl nodes \cite{Morimoto2016,Ahn2020a,Ma2021}. Indeed, experiments on non-centrosymmetric Weyl semimetals, including \WTe{} \cite{Wang2019a}, \TaIrTe{} \cite{Ma2019a,Shao2021}, MoTe$_2$ \cite{Ji2019} and TaAs \cite{Ma2017,Osterhoudt2019,Sirica2019,Gao2020}, have demonstrated signatures of shift and injection currents, which correspond to the components of the BPVE controlled by linearly and circularly polarized light, respectively. However, these observations rely on scanning photocurrent microscopy (SPCM) \cite{Ma2017,Wang2019a,Ma2019a,Shao2021,Ji2019,Osterhoudt2019} or terahertz emission \cite{Sirica2019,Gao2020}, neither of which can resolve the microscopic details of the photocurrent flow. In SPCM, a focused laser beam is rastered on a device while recording the total current between two distant contacts \cite{Graham2013}.  For gapless materials, the global diffusive photocurrent collected by SPCM is indirectly induced by the intrinsic photocurrent local to the photoexcitation through the Shockley-Ramo mechanism \cite{Song2014}. This local photocurrent is invisible to SPCM, yet captures the essential light-matter interaction.

\begin{figure*}[ht]
\includegraphics[scale=1]{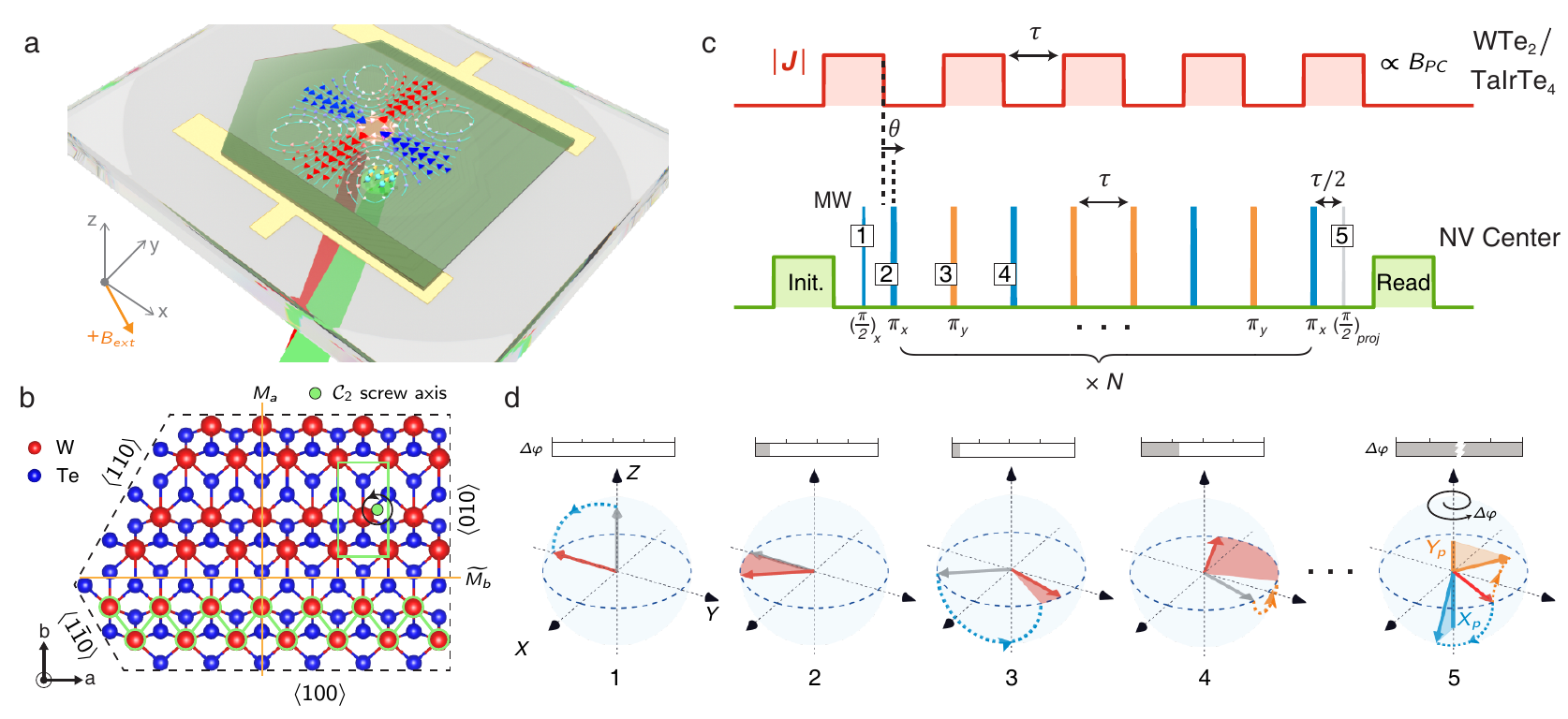}
\caption{\label{fig:1}Experimental overview. (a) Dual-beam scanning confocal microscope for PCFM and SPCM imaging. Samples are transferred onto prefabricated contacts on a diamond membrane containing a near-surface layer of NV centers. The linearly polarized red beam (661 nm) excites photocurrents in the sample, while the green beam (515 nm) interrogates the NV spins, which sense the local magnetic field. The direction of positive external magnetic field $B_{ext}$ is diagrammed. (b) Crystal structure of ${T_d}$-\WTe{} in the $ab$-plane. A single layer within the two-layer unit cell is shown. The unit cell, $C_2$ screw axis, and tungsten zigzag chain are highlighted in green. Mirror plane $M_a$ (glide mirror plane $\widetilde{M_b}$) exists perpendicular to the $a$-axis ($b$-axis). (c) Quantum lock-in detection of photocurrent. The green laser and first microwave (MW) pulse initializes a NV superposition state. Pulsed photocurrents $\boldsymbol{J}$ are generated by the red laser, which is chopped at the same rate as a series of MW $\pi$-pulses applied to the NV center. The MW sequence (XY8-$N$) allows only the resonant photocurrent magnetic field, with component $B_{PC}$ along the NV center axis, to affect the NV spin precession. The last green laser pulse reads out the final state to determine the total phase precession. We use an XY8-$2$ sequence with 16 $\pi$-pulses individually spaced by $\tau = $ 6.6~$\mu$s. (d) Bloch sphere illustrations of the instantaneous NV state (red arrow) at various points during the sensing sequence, as labeled in c). The phase difference $\Delta \varphi$ relative to evolution under zero ac magnetic field is shown in the bar graphs. The final projection pulse (`5') rotates the NV state around either the Bloch sphere's $X$ or $Y$-axis to probe the equatorial projections $X_P$ or $Y_P$.}
\end{figure*}

To clarify their photocurrent response, here we directly visualize the two-dimensional (2D) photocurrent flow in the type-II Weyl semimetals \WTe{} \cite{Soluyanov2015} and \TaIrTe{} \cite{Koepernik2016}. Our unique photocurrent flow microscopy (PCFM) technique is enabled by the high-sensitivity magnetic imaging of the photocurrent's Oersted field with nitrogen-vacancy (NV) centers in diamond \cite{Zhou2019}. We resolve for the first time the vector photocurrent density \boldjr{} at the point of photoexcitation and discover that it surprisingly circulates in material-distinctive patterns aligned with the Weyl semimetal's crystal lattice. This intimate visualization, explained by our theoretical and ab initio simulations, reveals that the local photocurrent is driven by an overlooked broken symmetry of the bulk: the anisotropy in the in-plane thermopower. Using simultaneous SPCM and PCFM imaging, we establish this anisotropic photothermoelectric effect (APTE), previously unknown, as the stimulus for the global photocurrent in our Weyl semimetal devices through the Shockley-Ramo theorem. Our observations prompt a careful review of the contribution of nonlinear shift currents to the strong edge and bulk photocurrent response in \WTe{} \cite{Wang2019a} and \TaIrTe{} \cite{Ma2019a,Shao2021}. Concomitantly, they open novel concepts for broadband, position-sensitive photodetectors using homogeneous materials with intrinsic Seebeck anisotropy.



Figure \ref{fig:1}a displays our room-temperature experimental configuration that extends Ref. \cite{Zhou2019}, which pioneered the detection, but not 2D imaging of photocurrents with NV ensemble magnetometry. Here, we use a thin diamond membrane to optically access non-transparent photocurrent samples and improve the NV photon collection efficiency for the challenging imaging measurements required for the model-free reconstruction of photocurrent flow. By transferring exfoliated flakes onto pre-fabricated electrodes on the diamond substrate, we are able to perform both PCFM and SPCM \emph{in-situ} on the same device. In PCFM, the NV center probe beam (green, 515 nm) is scanned around a fixed position for the photocurrent excitation beam (red, 661 nm) to map the local magnetic field. Alternatively, by measuring the total collected current while rastering either beam, we can acquire SPCM images at either wavelength.

\begin{figure*}[ht]
\includegraphics[scale=1]{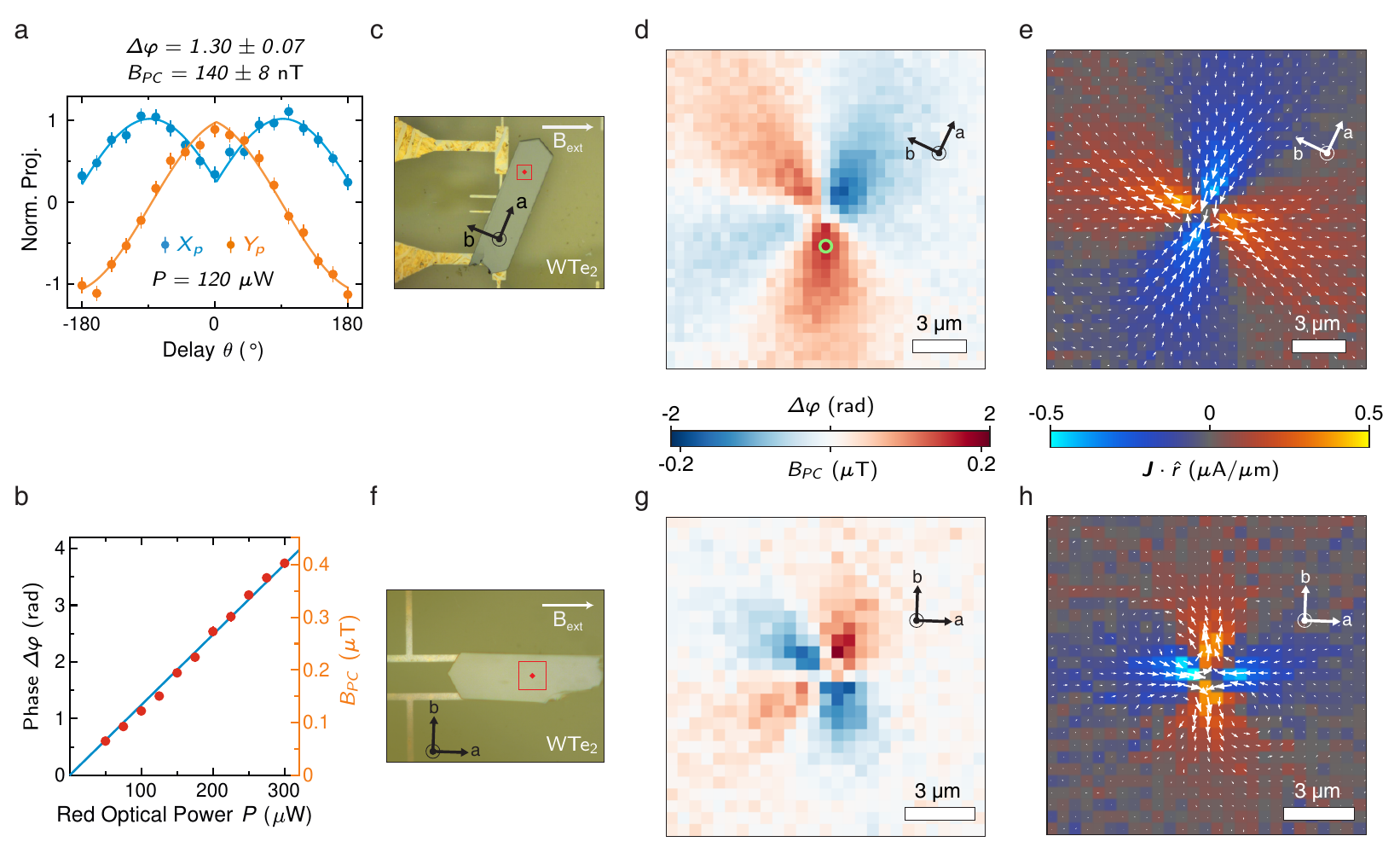}
\caption{\label{fig:2}Crystal-axes-aligned bulk photocurrents in \WTe. (a) The $X_P$ and $Y_P$ projections of the final NV state versus the delay $\theta$ between the photocurrent and the NV spin-driving pulses. The probed NV position is the green circle in d). We extract a maximal NV phase precession $\Delta \varphi$ = 1.30 $\pm$ 0.07 rad, corresponding to a magnetic field $B_{PC}$ = 140 $\pm$ 8 nT along the NV center axis, for a photocurrent excitation (red) power $P$ = 120 $\mu$W. (b) Precession angle $\Delta \varphi$ versus optical power $P$. The linear dependence of $\Delta \varphi$ ($\propto |\boldsymbol{J}|$) on $P$ is consistent with electrical measurements of the global photocurrent $I_{global}$. (c) Optical micrograph of \WTe{} Device A, with sample thickness 1.8 $\mu$m. The long edges identify the $a$-axis of \WTe. The in-plane projection of $B_{ext}$ (= 28 mT) is denoted by the white arrow, but has no influence on the results. (d) Spatial image of the NV center precession angle $\Delta \varphi(\boldsymbol{r}) \propto B_{PC}(\boldsymbol{r})$ in the interior of \WTe{} Device A, with the photoexcitation fixed at the center of the image ($P$ = 120 $\mu$W). The imaged region is the red box in c). (e) The photocurrent density \boldjr{} reconstructed from $B_{PC}(\boldsymbol{r})$ in d) as a 2D vector field. The radial component of the flow ($\boldsymbol{J} \cdot \hat{\boldsymbol{r}}$) is superimposed as a false colormap. The local photocurrent circulates by flowing in along the $a$-axis and out along the $b$-axis. (f) Optical micrograph of \WTe{} Device B, with sample thickness 360 nm. (g) Spatial image of $B_{PC}(\boldsymbol{r})$ for the boxed region in f) with $P$ = 120 $\mu$W. (h) Reconstructed \boldjr{} in Device B. The four-fold photocurrent circulation remains aligned with the crystal axes of the rotated flake.}
\end{figure*}

The type-II Weyl semimetals \WTe{} and \TaIrTe{} crystallize in an inversion-breaking orthorhobmic structure (space group $Pmn2_1$). As shown in Fig. \ref{fig:1}b for a single layer of $T_d$-\WTe{}, the transition metal atoms form quasi-one-dimensional zigzag chains along the $a$-axis within the \mbox{$ab$-plane}, leading to anisotropic electrical, optical, and thermal properties \cite{Kang2019,Frenzel2017,Chen2019b}. The lowest-order shift photocurrent is described by $J_i = \Re(\sigma^{(2)}_{ijk}) E_j(\omega)E_k^{\text{*}}(\omega)$, where $\Re(\sigma^{(2)}_{ijk})$ is the real part of the second order susceptibility tensor and $E_j(\omega) = E_{0,j} e^{i\omega t}$ are the Cartesian components of the light's electric field \cite{Sipe2000,Tan2016,Morimoto2016,Ma2021}. For a system with inversion symmetry, $\sigma^{(2)}_{ijk}$ must identically vanish since $E_j(\omega)E_k^{\text{*}}(\omega)$ is even under inversion, while $J_i$ is odd. Moreover, for light at normal incidence, a shift current in the $ab$-plane of \WTe{} or \TaIrTe{} is prohibited by a similar argument using the two-fold rotation symmetry ($C_2$ screw axis) within the $ab$-plane (Fig. \ref{fig:1}b) \cite{Wang2019a,Ma2019a,Shao2021}. 


In SPCM experiments on \WTe{} \cite{Wang2019a} and \TaIrTe{} \cite{Ma2019a,Shao2021}, photocurrents were detected when exciting mirror-symmetry-breaking edges and within the sample's bulk near narrow electrical contacts. The second-order shift current can in principle exist at the edges due to their symmetry-breaking effect, but the interior photocurrents require additional justification. Ref.~\cite{Ma2019a} interpreted the bulk response in \TaIrTe{} as a third-order process $J_i = \sigma^{(3)}_{ijkl} E^{DC}_j E_k(\omega)E_l^{\text{*}}(\omega)$, where the normally incident light field mixes with an uncontrolled dc electric field $E^{DC}_j$ that arises from device-specific interfaces or inhomogeneities. Alternatively, in the near-field SPCM experiment of Ref. \cite{Shao2021}, the metallic tip transduces the incident light into an additional out-of-plane electric field ($E_c(\omega)$ for \mbox{$ab$-plane} samples). The authors argued that their bulk signal arises from non-planar second-order susceptibility elements (e.g., $\sigma_{aac}$ and $\sigma_{bbc}$) that are symmetry-allowed, but inaccessible for normally incident light. Although such mechanisms may contribute, we reveal that the uncovered APTE is an unified cause of bulk and edge photocurrents in \WTe{} and \TaIrTe{}.


To achieve the enhanced sensitivity to detect sub-$\mu$A/$\mu$m photocurrent densities, we utilize a multi-pulse quantum lock-in sequence for ac magnetic fields \cite{Zhou2019,Ku2020,Zhang2021,Vool2021}. As shown in Fig. \ref{fig:1}c, we strobe the photocurrent excitation on the sample and simultaneously apply a sequence of microwave $\pi$-pulses (\mbox{XY8-$N$}) to manipulate the NV center spin \cite{Zhou2019}. When the $\pi$-pulses are spaced at the same interval $\tau$ as the on/off photocurrent control, they isolate the effect of the photocurrent's magnetic field on the NV spin precession, while filtering out noise and significantly extending the NV center's coherence time ($>$150~$\mu$s). Figure \ref{fig:1}d diagrams the evolution of the NV superposition state $\ket{\psi}$ on the Bloch sphere at various times in a measurement sequence with timing delay $\theta \approx 20^\circ$ between the photocurrent and microwave pulses. The final state is rotated by a $\pi/2$-pulse around either the $X$- or $Y$-axis of the Bloch sphere and then read out by NV fluorescence to yield the projections $X_P$ or $Y_P$, from which the state's precession angle $\Delta\varphi$ is determined as $\Delta\varphi = \arctan(Y_P/X_P)$ (Methods).

We first demonstrate photocurrent detection for linearly polarized light in the interior of a \WTe{} sample (Device A). Figure \ref{fig:2}a measures the NV spin precession versus the delay $\theta$ for a fixed spatial offset (2 $\mu$m vertically) between the photocurrent and NV probe beams. Fitting the oscillations in $X_P(\theta)$ and $Y_P(\theta)$ simultaneously \cite{Zhou2019}, we deduce that a maximal phase precession $\Delta \varphi = 1.30 \pm 0.07$ is obtained at $\theta = 0^\circ$ for a photocurrent excitation power $P = 120~\mu$W. The photocurrent's local magnetic field $B_{PC}$ along the NV center axis is then determined as $B_{PC} = \Delta\varphi / (2\pi \gamma_e \times  4 N \tau) = 140 \pm 8$~nT, where $\gamma_e = 28$~kHz/$\mu$T is the electron gyromagnetic ratio and the photocurrent pulses are well approximated as square wave (Supplementary Fig. 1b).

\begin{figure*}[t]
\includegraphics[scale=1]{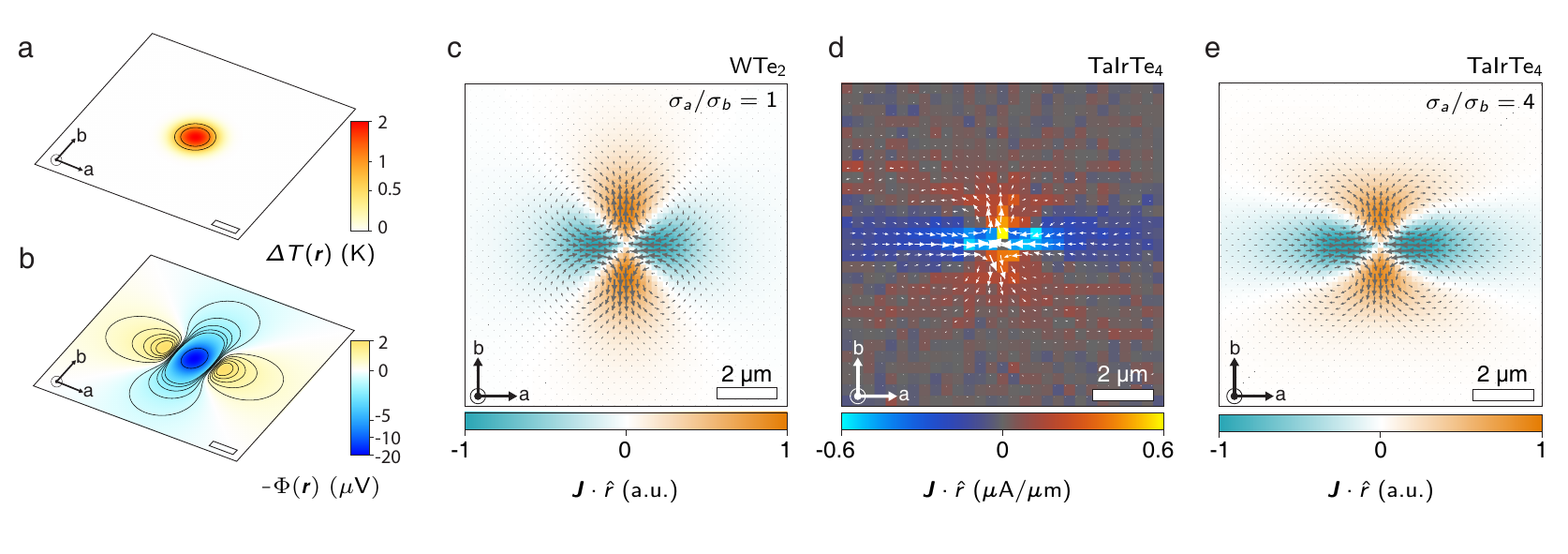}
\caption{\label{fig:3}APTE in anisotropic Weyl semimetals. (a) Simulated temperature rise $\Delta T(\boldsymbol{r})$ in a \WTe{} flake for $P$ = 120 $\mu$W. The temperature distribution, which is nearly isotropic, drives the majority electron carriers radially outward. (b) Electrochemical potential $\Phi(\boldsymbol{r})$ obtained by solving Eq. \ref{eq:1} using a modeled Seebeck anisotropy $|S_a - S_b|$= 10 $\mu$V/K and a conductivity anisotropy $\sigma_a/\sigma_b$ = 1. The electromotive force on electrons points toward lower values of $-\Phi(\boldsymbol{r})$ (blue). The scalebars in (a-b) denote 2~$\mu$m. (c) Simulated interior photocurrent distribution \boldjr{} for \WTe{}, comprising the sum of the Seebeck and diffusion terms represented in a) and b). Electron carriers (opposite of the conventional current) undergo a four-fold circulation from the axis with the more negative Seebeck coefficient ($a$-axis) to the axis with the less negative Seebeck coefficient ($b$-axis). (d) Experimentally imaged \boldjr{} in the interior of \TaIrTe{} Device A ($P = 90~\mu$W). The photocurrent density in \TaIrTe{} is more anisotropic than in \WTe, but obeys the same directionality relative to the orthorhombic lattice. (e) Simulated \boldjr{} for \TaIrTe{} within the APTE model using a conductivity anisotropy $\sigma_a/\sigma_b$ = 4.}
\end{figure*}

By varying $P$, we find that $\Delta\varphi$, and by extension the local photocurrent density $|\boldsymbol{J}|$, is linear in excitation powers up to 300 $\mu$W (Fig. \ref{fig:2}b). The linear power dependence for $|\boldsymbol{J}|$ is consistent with the behavior of the global photocurrent ($I_{global}$) that we will electrically detect by SPCM when directing the photoexcitation at symmetry-breaking edges (Extended Data Fig. \ref{ext:1}h). Although $I_{global} \propto P$ has been taken to support a shift current mechanism that is second order in light's electric field ($\propto E_j E_k$) \cite{Wang2019a}, it actually reflects here a non-saturating PTE due to high thermal conductance.


In Fig. \ref{fig:2}d, we image the spatially-varying phase precession $\Delta\varphi(\boldsymbol{r})$ in \WTe{} Device A, shown in Fig. \ref{fig:2}c, by scanning the NV center probe beam relative to the photocurrent excitation at the center of the image. An unexpected four-fold, sign-switching pattern is observed, signaling intricate photocurrent flow. By inverting the Biot-Savart equation with the steady state condition $\nabla \cdot \boldsymbol{J} = 0$, we can uniquely reconstruct the 2D photocurrent distribution ($J_x, J_y$) from its magnetic field component $B_{PC}(\boldsymbol{r}) \propto \Delta\varphi(\boldsymbol{r})$ \cite{Chang2017,Tetienne2016,Ku2020,Rohner2018} (see Supplementary Section 2). Exfoliated \WTe{} exhibits long edges parallel to the crystallographic $\langle 100 \rangle$ direction ($a$-axis) and shorter diagonal edges parallel to the $\langle 110 \rangle$ directions (see Fig. \ref{fig:2}c and schematic in Fig. \ref{fig:1}b). Remarkably, the reconstructed current in Fig. \ref{fig:2}e displays a $C_{2v}$ symmetry that is aligned to the crystallographic axes. The conventional current (positive charge) flows predominantly inward along the $a$-axis and outward along the $b$-axis, as visualized by plotting $\boldsymbol{J} \cdot \hat{\boldsymbol{r}}$ in false color, where $\hat{\boldsymbol{r}}$ is the normalized vector from the photoexcitation origin.

We verify that the direction of the external magnetic field $B_{ext}$ has no bearing on our observations (see Extended Data Fig. \ref{ext:2} for absence of photo-Nernst effect). Moreover, we measure a second device (Device B) with only the \WTe{} flake rotated, while keeping the same orientation for $B_{ext}$ and the NV center axis (Fig. \ref{fig:2}f). The phase map $\Delta\varphi(\boldsymbol{r})$ accordingly rotates, with a slight change in shape since the NV center now probes a different component of the field relative to the current distribution. The reconstructed photocurrent flow, however, is qualitatively the same, remaining aligned to the $a/b$-axis of the rotated sample (Fig. \ref{fig:2}h).

As we have no out-of-plane $E_c$ and any accidental $E^{DC}_j$ is unlikely to be aligned with the crystal axes, we consider whether a non-shift current mechanism induces the bulk photocurrent. Supposing different electrical conductivities $\sigma_{a/b}$ and Seebeck coefficients $S_{a/b}$ along the $a/b$-axis of the \WTe{} crystal, which are aligned along the lab $x/y$-axis, the photocurrent density can be modeled as
\begin{equation}
\begin{split}\label{eq:1}
	J_x(\boldsymbol{r}) &= -\sigma_{a}\big(\partial_{x} \Phi(\boldsymbol{r}) + S_{a}\cdot\partial_{x} T(\boldsymbol{r}) \big)\\
	J_y(\boldsymbol{r}) &= -\sigma_{b}\big(\partial_{y} \Phi(\boldsymbol{r}) + S_{b}\cdot\partial_{y} T(\boldsymbol{r}) \big).
\end{split}
\end{equation}
Here, the total photocurrent $\boldsymbol{J}$ is contributed by a diffusion term $\boldsymbol{J}_{d} = -\boldsymbol{\sigma} \nabla\Phi$ due to an induced electrochemical potential $\Phi$ and a Seebeck term $\boldsymbol{J}_{ph} = -\boldsymbol{\sigma}\boldsymbol{S}\nabla T$ due to the temperature profile $T$, with $\boldsymbol{\sigma}$ and $\boldsymbol{S}$ as the conductivity and Seebeck tensors. Given a nonuniform $T(\boldsymbol{r})$, the photocurrent $\boldsymbol{J}(\boldsymbol{r})$ is determined by solving for the steady state $\Phi(\boldsymbol{r})$ that satisfies the continuity ($\nabla \cdot \boldsymbol{J} = 0)$ and boundary conditions on $\boldsymbol{J}$. Our analysis reveals that a circulating $\boldsymbol{J}$ emerges only when the thermopower is anisotropic, with amplitude $|\boldsymbol{J}| \propto |S_a-S_b|$ (see Methods). On the other hand, anisotropy in the conductivity $\sigma_{a/b}$ and temperature distribution $T(\boldsymbol{r})$ determine the shape of $\boldsymbol{J}(\boldsymbol{r})$, but cannot cause nonzero current. Similar transport equations have been considered for anisotropic thermoelectrics \cite{Lukosz1964}; however, to our knowledge, the predicted spontaneous currents have never been resolved due to challenging requirements on magnetic field and spatial resolution.

Figure \ref{fig:3}a displays the simulated temperature rise $\Delta T(\boldsymbol{r})$ under laser heating in a \WTe{} flake for an illustrative 2D thermal model (see Supplementary Fig. 1 for full 3D thermal simulations). At room temperature, the Seebeck coefficient for \WTe{} is negative \cite{Kabashima1966,Wu2015a,Jana2015,Rana2018}, which implies that net electron carriers are driven radially outward by the gradient $\nabla T$. In an isotropic material, the resulting $\Phi$ exactly counterbalances the Seebeck flow, yielding zero PTE photocurrents in the bulk. However, if $S_a \neq S_b$, our simulations reveal that electrons flow outward along the axis with the more negative Seebeck coefficient and return inward along the orthogonal direction in steady state.


\begin{figure*}[t]
\includegraphics[scale=1]{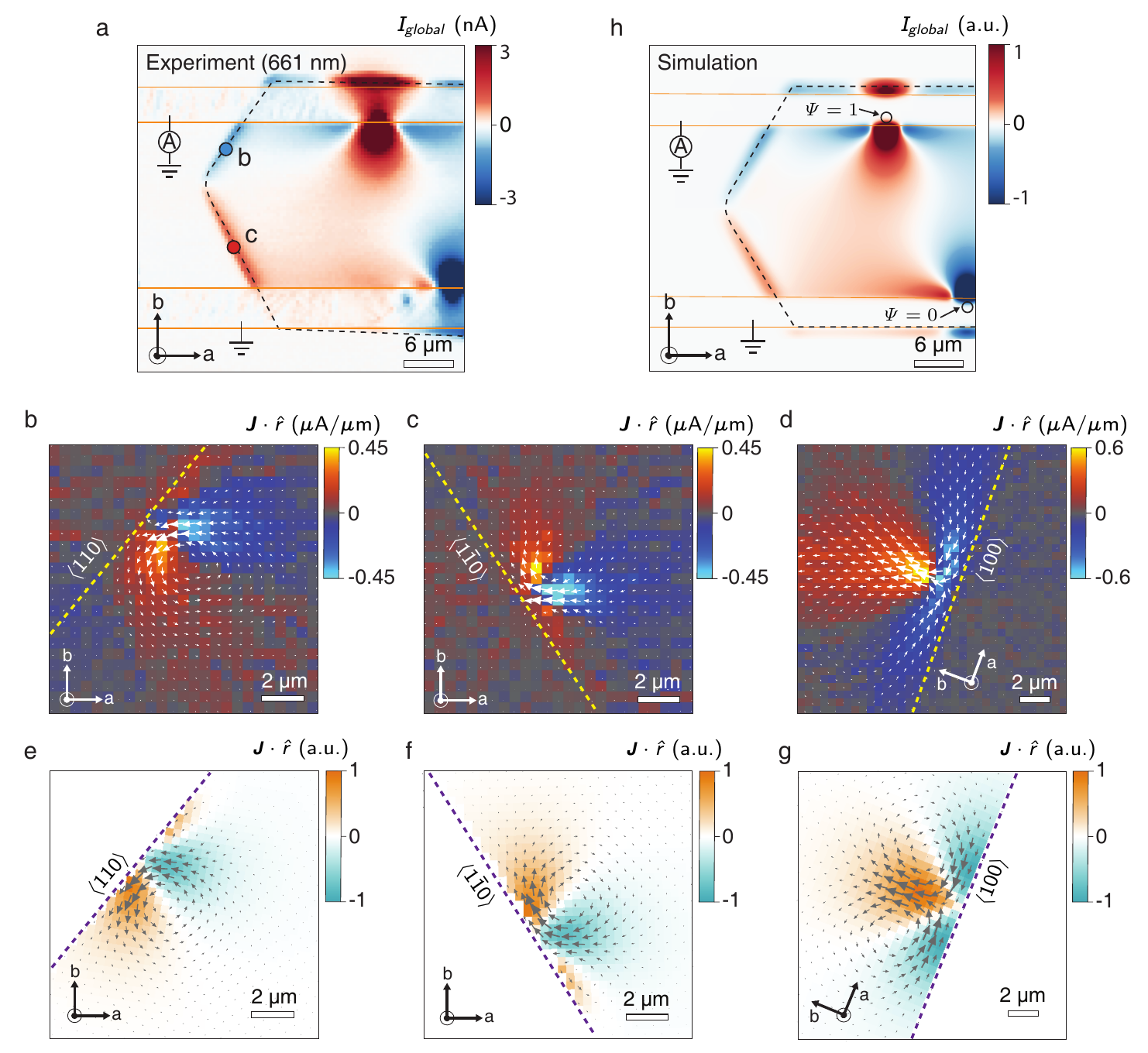}
\caption{\label{fig:4}APTE as the stimulus for long-range photocurrent in \WTe{} devices. (a) Experimental SPCM image of $I_{global}$ in \WTe{} Device B for $P$ = 300 $\mu$W at 661 nm. Robust edge photocurrents are detected along the $\langle 110 \rangle$ and $\langle 1\bar{1}0 \rangle$ edges, with a negative (blue) and positive (red) polarity, respectively. Sign-switching interior photocurrents are also collected relative to two localized points on the upper and lower gold pads (orange). (b) Simultaneous PCFM image of $\boldsymbol{J}(\boldsymbol{r})$ for photoexciting the $\langle 110 \rangle$ edge in Device B. (c) PCFM image for photoexciting the $\langle 1\bar{1}0 \rangle$ edge. The photoexcitation power for b) and c) is 120~$\mu$W, and their locations are labeled in a). The polarity of the edge photocurrent $I_{global}$ is immediately visualized by the net component of $\boldsymbol{J}(\boldsymbol{r})$ parallel to the edge. (d) PCFM image for photoexciting the $\langle 100 \rangle$ edge in \WTe{} Device A with $P = 100~\mu$W. (e,f,g) Simulated $\boldsymbol{J}(\boldsymbol{r})$ for the $\langle 110 \rangle$, $\langle 1\bar{1}0 \rangle$, and $\langle 100 \rangle$ edges, respectively, using the APTE model. (h) Simulation of the full SPCM image under Shockley-Ramo theory. The simulation assumes that $\boldsymbol{J}_{ph}(\boldsymbol{r})$ flows according to anisotropic Seebeck coefficients and that the weighting field $\nabla \Psi(\boldsymbol{r})$ is determined by the boundary conditions $\Psi$ = 1, 0 on two point-like surfaces labeled on the upper and lower pads, respectively.}
\end{figure*}

Thus, our imaging (Fig. \ref{fig:2}e,h) identifies the $a$-axis as possessing the more negative Seebeck coefficient ($S_a < S_b$). Assuming $|S_a-S_b| = 10~\mu$V/K and an approximately isotropic conductivity $\sigma_a/\sigma_b$ = 1 (see Supplementary Fig. 3), we show in Fig. \ref{fig:3}b the equilibrium potential $-\Phi(\boldsymbol{r})$ corresponding to $\Delta T(\boldsymbol{r})$. Since the electromotive force on electrons points towards smaller values (blue) of $-\Phi(\boldsymbol{r})$, Fig. \ref{fig:3}b illustrates how $\Phi$ drives electron backflow opposite to $\nabla T$, as well as electron circulation from the $a$-axis to the \mbox{$b$-axis}. In Fig. \ref{fig:3}c, we calculate the full photocurrent distribution $\boldsymbol{J}(\boldsymbol{r})$ using Eq. \ref{eq:1}, showing excellent agreement with our \WTe{} images (Figs. \ref{fig:2}e,h). A quantitative estimate of the Seebeck anisotropy is complicated by the dependence of $|\boldsymbol{J}|$ on not only $|S_a-S_b|$, but also on the conductivities $\sigma_{a/b}$ and the internal temperature $\Delta T$ that cannot be directly measured. Estimating these factors, we describe our best experimental determination $|S_a-S_b| = 8^{+8}_{-5}~\mu$V/K in Supplementary Section 4.

Notably, our density functional calculations also predict negative Seebeck coefficients $S_a$ and $S_b$ for \WTe{} at room temperature, with $S_a < S_b$ in agreement with experiment (Methods). For semimetals, both electron and hole bands contribute to the thermopower, with opposite signs and a weighting factor determined by the band's density of states and squared group velocity in the direction of transport. As shown in Extended Data Fig. \ref{ext:3}d, the predicted $S_a$ and $S_b$ diverge above 200 K, with $S_a$ becoming more negative and $S_b$ more positive, leading to $|S_a - S_b| \approx 15~\mu$V/K at 300 K. This trend is mainly contributed by the anisotropy in the electron velocities for the low-lying conduction bands. As temperature increases, the thermopower senses the larger density of states for holes away from the Fermi level, but this positive contribution is outweighed by the electron contribution along the $a$-axis, where the electron velocities are highest.

To establish the generality of the APTE, we perform PCFM imaging on \TaIrTe{}, where the $a$-axis zigzag chains are composed of alternating Ta and Ir atoms \cite{Koepernik2016}. \TaIrTe{} also possesses a negative Seebeck coefficient at room temperature \cite{LeMardele2020}, but exhibits more anisotropic in-plane electrical transport than \WTe{}, with higher conductivity along the $a$-axis \cite{Liu2018b,Kumar2021}. Our PCFM image of \TaIrTe{} shows that photoexcited electrons again flow out along the $a$-axis and in along the $b$-axis (Fig. \ref{fig:3}d). However, a stronger asymmetry is observed, where the flow remains parallel to the $a$-axis longer than it does to the $b$-axis. This distinct shape for $\boldsymbol{J}(\boldsymbol{r})$ is reproduced in our APTE model (Fig. \ref{fig:3}e) by using a higher conductivity anisotropy $\sigma_a/\sigma_b \approx 4$ (Supplementary Section 3), which is consistent with prior determinations \cite{Kumar2021,Shao2021}. 

We now connect the APTE to the global photocurrent collected in our Weyl semimetal devices. In gapless materials, photocurrent collection can be mapped onto the Shockley-Ramo theorem, providing the insight that the charge flowing into a distant contact ($I_{global}$) is dominated by the induced diffusion current $\boldsymbol{J}_{d}$, rather than by the directly photoexcited $\boldsymbol{J}_{ph}$, which remains local to the illumination. Song and Levitov \cite{Song2014} showed that $I_{global}$ can be simply calculated as
\begin{equation}\label{eq:2}
	I_{global} = C \int \boldsymbol{J}_{ph}(x,y) \cdot \nabla \Psi(x,y) dx dy,
\end{equation}
using only the direct photocurrent $\boldsymbol{J}_{ph}(x,y)$, here due to the anisotropic Seebeck flow, and an auxiliary weighting potential $\Psi(x,y)$ determined by solving Laplace's equation within the particular device geometry (Supplementary Section 6). The constant $C$ depends on circuit resistances. Thus, nonzero $I_{global}$ can generally arise if the areal contributions of $\boldsymbol{J}_{ph} \cdot \nabla \Psi$ to the integral do not cancel due to differing symmetries for $\boldsymbol{J}_{ph}$ and $\nabla \Psi$.




Figure \ref{fig:4}a displays our SPCM image at 661 nm excitation for \WTe{} Device B (Fig. \ref{fig:2}f). Long-range photocurrents are detected along the entirety of the oblique $\langle 110 \rangle$ and $\langle 1\bar{1}0 \rangle$ edges, with a negative and positive polarity, respectively. In addition, we collect $I_{global}$ in the interior of the device, with a sign-switching pattern surrounding two hot spots along the upper and lower contacts. These spatial patterns reproduce the salient features, attributed to nonlinear responses, in prior SPCM measurements \cite{Ma2019a,Wang2019a,Shao2021}. Particularly, global photocurrents were collected in \WTe{} \cite{Wang2019a} and \TaIrTe{} \cite{Shao2021} when exciting oblique (low-symmetry) edges, but not high-symmetry $\langle 100 \rangle$ or $\langle 010 \rangle$ edges. Moreover, Refs. \cite{Ma2019a,Shao2021} observed $I_{global}$ to switch sign near narrow contacts, depending on whether the nearest edge of the contact was parallel to the $a$-axis or $b$-axis.


To reveal the microscopic stimulus for our global photoresponse, we switch to PCFM imaging on the same device with the photocurrent laser fixed at $I_{global}$ hot spots. Figure \ref{fig:4}b and \ref{fig:4}c display the photocurrent flow when exciting the upper $\langle 110 \rangle$ and lower $\langle 1\bar{1}0 \rangle$ edges, respectively. The local photocurrent distribution appears as though the edge truncates the four-fold APTE pattern observed in the interior of the \WTe{} device (Fig. \ref{fig:2}e,h). This is corroborated in Fig. \ref{fig:4}d, where we image the photocurrent flow when exciting a $\langle 100 \rangle$ edge on \WTe{} Device A (Fig. \ref{fig:2}c). The immediate visual interpretation of these images explains the presence or absence, as well as the polarity, of the global edge photocurrent. For photoresponsive oblique edges (Figs. \ref{fig:4}b,c), \boldjr{} possesses a strong component parallel to the edge, with direction set by the truncation of the ``$a$-axis in, $b$-axis out'' APTE flow. For nonresponsive $\langle 100 \rangle$ or $\langle 010 \rangle$ edges (Fig. \ref{fig:4}d), the component of \boldjr{} parallel to the edge cancels, as consistent with the mirror symmetry about the perpendicular to these edges.


More precisely, NV magnetometry senses the total photocurrent \boldjr{} contributed by both the Seebeck current $\boldsymbol{J}_{ph}$ and the diffusion current $\boldsymbol{J}_{d}$. For the APTE, $\boldsymbol{J}_{d}$ possesses both a strong local (circulating) component and a weaker long-range component, where the latter can be extracted by the contact due to its equipotential boundary condition. Our measured $I_{global}$, reflecting mainly the long-range component of $\boldsymbol{J}_{d}$, is only a few nanoamps. Meanwhile, the simultaneously imaged total current density \boldjr{} near the edge is $\sim$100 nA/$\mu$m, indicating that the source/drain effect of the contacts is a small perturbation to the local APTE pattern. Hence, we compare our data (Fig. \ref{fig:4}b,c,d) to local simulations of $\boldsymbol{J}(\boldsymbol{r})$ due to the APTE (Eq.~\ref{eq:1}) without considering the distant contacts (see Supplementary Fig. 7 for simulations over the whole device geometry including contacts).  Figures \ref{fig:4}e,f,g present the simulated APTE patterns that  incorporate the modified temperature distribution and parallel flow boundary condition near the edge for the $\langle 110 \rangle$, $\langle 1\bar{1}0 \rangle$ and $\langle 100 \rangle$ edges, respectively, showing excellent agreement with experiment (Figs. \ref{fig:4}b,c,d).

Continuing, we simulate the full SPCM image via the Shockley-Ramo theorem (Eq. \ref{eq:2}). We approximate $\boldsymbol{J}_{ph}(\boldsymbol{r,r_0})$ for each photoexcitation location $\boldsymbol{r_0}$ by assuming that the temperature distribution $T(\boldsymbol{r},\boldsymbol{r_0})$ is cropped by the device boundaries if $\boldsymbol{r_0}$ is proximal to the edge. The details of $I_{global}(\boldsymbol{r_0})$ also depend on $\nabla\Psi(\boldsymbol{r})$, which is determined by how electrical contact is made to the device. In \WTe{} Device B, our modeling (Fig. \ref{fig:4}h) reveals that the experimental $I_{global}$ can be reproduced by a simple point-like contact (white circle) in each of the upper and lower leads. Evidently, although our bottom contacts appear extended, they form effective point-like conductive interfaces in this device. We confirm that $\boldsymbol{J}(\boldsymbol{r})$ near the contacts (Extended Data Fig. \ref{ext:1}i) remains virtually identical to that in the interior of the device (Fig. \ref{fig:2}h). Thus, although $\boldsymbol{J}_{ph}(\boldsymbol{r})$ is $C_{2v}$ symmetric in the bulk, nonzero $I_{global}$ can still be collected for non-uniform $\nabla \Psi$. Our bulk photocurrent thus stems from the interaction between $\boldsymbol{J}_{ph}(\boldsymbol{r})$ and the changing direction and strength of $\nabla \Psi$ as the point-like contacts are approached (Supplementary Fig. 8). Finally, we observe qualitatively identical long-range edge and interior photocurrents in \TaIrTe{} devices (Extended Data Fig. \ref{ext:4}), establishing the APTE as a ubiquitous and primary mechanism in anisotropic Weyl materials.




\emph{Discussion.---} The APTE is distinct from photocurrent generation mechanisms based on interfaces \cite{Sunku2020,Akamatsu2021,Jiang2021,Gabor2011,Buscema2013,Lee2014,Deng2014} or surface states \cite{McIver2012} and from rotationally symmetric mechanisms, such as the photo-Nernst effect \cite{Cao2016,Zhou2019} or unbalanced electron-hole diffusion \cite{Liu2015a,Ma2019}, which all do not stimulate collectable long-range photocurrents within the homogeneous bulk. Moreover, unlike the BPVE, whose response is peaked for certain photon energies \cite{Tan2016,Patankar2018}, the APTE is effective for broad wavelength bands through the material's optical absorption and may appear in a wide range of crystal structures. In Extended Data Fig. \ref{ext:5}, we propose prototype device designs based on the APTE that enhance the collection efficiency for bulk photocurrents and enable four-quadrant, position-sensitive photodetection in a single chip at room temperature.


Our visualization of spontaneous vortical currents in thermoelectrically anisotropic materials may also be relevant to experiments on similar materials where temperature gradients could arise, for example, by resistive heating. An intriguing extension is to explore photocurrent in the hydrodynamic regime of \WTe{} \cite{Vool2021,Aharon-Steinberg2022} to discern the effects of electronic shear viscosity on the APTE-driven current circulation appearing here in the diffusive limit. Moreover, our dual-beam quantum magnetometry technique opens the non-contact measurement of thermoelectric effects in anisotropic superconductors, where the interplay between the quasiparticle current and a supercurrent counterflow may engender novel spatial distributions \cite{Ginzburg1991}.





\clearpage{}
\section{Acknowledgments}
The authors thank Q. Ma, M. Jung, and P. C. Jerger for valuable discussions. B.B.Z. acknowledges support from the National Science Foundation (NSF) under CAREER award number DMR-2047214 and award number ECCS-2041779. This material is based on work supported by the Air Force Office of Scientific Research under award numbers FA2386-21-1-4095 and FA2386-21-1-4059. Y.R. acknowledges support from the NSF award number DMR-1712128. C.L. and D.B. were supported by the U.S. Department of Energy, Office of Science, Basic Energy Sciences under award number DE-SC0021071 (ab initio calculations of band structure and Seebeck coefficients of \WTe{}). Z.L. acknowledges supports from the Singapore National Research Foundation-Competitive Research Program under awards NRF-CRP22-2019-0007 and NRF-CRP21-2018-0007. This work was performed, in part, at the Integrated Sciences Cleanroom and Nanofabrication Facility at Boston College and at the Center for Nanoscale Systems (CNS), a member of the National Nanotechnology Infrastructure Network, which is supported by the NSF under award number ECCS-0335765. CNS is part of Harvard University.

%

\section{Methods}
\subsection{Sample details}
High-quality crystals of \WTe{} are grown in a multi-step process. First, a powder specimen of \WTe{} is synthesized by heating a stoichiometric mixture of W and Te (both with 99.999\% purity) inside an evacuated silica tube. The mixture is heated to 450 $^\circ$C for 24 hours and 800 $^\circ$C for another 24 hours at the rate 1 $^\circ$C/min, followed by quenching in water. Second, the mixture of powder \WTe{} (0.2 g) and Te (10 g) is heated to 825 $^\circ$C at 2 $^\circ$C/min, held for 2 days, cooled to 525 $^\circ$C at 4~$^\circ$C/hour, and centrifuged to remove the excess Te flux. Third, we anneal the flux-grown samples under vacuum in a two-zone furnace with hot end at 415 $^\circ$C and cold end at 200 $^\circ$C for two days to remove any remaining Te impurity and reduce defects.

Bulk \TaIrTe{} crystals are synthesized via the self-flux method. A total of 2.0 g powder with the molar ratio of Ta:Ir:Te=1:1:12 are loaded in a silica tube, sealed under a high vacuum condition ($<10^{-2}$ Pa). Then, the silica tube is heated to 1000 $^\circ$C within three days and held at the temperature for seven days. Finally, the tube is cooled down to 600 $^\circ$C within 21 days. Needle-shaped \TaIrTe{} crystals are picked out from the ingot. 

\WTe{} and \TaIrTe{} flakes are exfoliated onto a polydimethylsiloxane (PDMS) stamp, which is then pressed onto and slowly peeled away from the diamond membrane to transfer targeted flakes. The diamond membrane used in this work is cut and polished to 100 $\mu$m thickness (Delaware Diamond Knives) from a bulk electronic grade diamond sample grown by chemical vapor deposition (Element Six). An NV ensemble is created in the diamond membrane by $^{15}$N ion implantation at 45 keV energy and an areal dose of $10^{12}$ ions/cm$^2$, followed by annealing at 1050 $^{\circ}$C for two hours. The XY8-1 coherence time for the sample is about 150 $\mu$s. Standard optical photolithography is used to pre-pattern several sets of Cr/Au bottom contacts onto the diamond membrane for SPCM measurements. 

\subsection{NV center measurements}
The diamond membrane with photocurrent sample facing up (Fig. \ref{fig:1}a) is mounted onto a quartz slide. The quartz slide features a lithographically defined coplanar waveguide to deliver the microwave pulses for NV center spin manipulation. The NV readout (515 nm) and photocurrent (661 nm) lasers, both linearly polarized, pass through the quartz slide and diamond membrane to impinge on the NV ensemble layer and adjacent photocurrent sample at the top diamond surface. PCFM imaging is performed with 370 $\mu$W green laser power and a dc photon count rate of $5 \cdot 10^6$ counts/s for the NV ensemble. As an example, the PCFM image in Fig. \ref{fig:2}d acquires 35 by 35 pixels with 500 nm resolution and 67 s integration time per pixel, for a total frame time of 23 hours.


To determine $\Delta \varphi$ at each pixel, we perform an XY8-2 pulse sequence with total NV center free precession time $16\tau$ = 105.6 $\mu$s, delayed by $\theta = 20^\circ$ relative to the pulsed photoexcitation \cite{Zhou2019}. The total photon integration time at each pixel is divided between four different final projection pulses: $X_{\pm \pi/2}$ and $Y_{\pm \pi/2}$, where, for example, $X_{+\pi/2}$ denotes a rotation of the NV superposition state $\ket{\psi}$ around the $X$-axis of the Bloch sphere by $+\pi/2$ radians. After each projection pulse, the NV photon count rate ($PL$) during the green laser readout pulse is recorded, and we determine the differential projections $X_P =  PL(X_{-\pi/2}) - PL(X_{+\pi/2})$ and $Y_P = PL(Y_{-\pi/2}) - PL(Y_{+\pi/2})$ to reject common-mode noise. The accumulated phase $\Delta \varphi$ with respect to the initial NV state $\ket{\psi_0} = (\ket{0} - i \ket{-1})/ \sqrt{2}$ prepared by an $X_{+\pi/2}$ pulse is then computed as $\Delta \varphi = \arctan(Y_P/X_P)$.

To avoid phase unwrapping ambiguities, we set the photocurrent excitation power $P$ such that the range of $\Delta \varphi$ in the PCFM image is always between $\pm \pi$. In addition, we utilize a small delay $\theta = 20^\circ$ between the microwave spin driving and photocurrent excitation pulses to mitigate the effect of phase broadening over the probed NV center spot size. If the photocurrent magnetic field $B_{PC}$ is nonuniform over the NV spot size, the coherence of the final NV state $\ket{\psi}$ is damped by a factor $\exp(-\sigma_{\Delta \varphi} ^2 (\Delta \varphi(\theta)/\Delta \varphi(0))^2 /2)$, where $\sigma_{\Delta \varphi}^2$ is the variance in the phase precession angle over the probed NV spot and $\Delta \varphi(\theta)$ is the mean precession angle for delay $\theta$ \cite{Zhou2019}. Optimizing the sensitivity thus requires balancing the amplitude of the phase precession $\Delta \varphi(\theta)$ (which decreases for larger $\theta$; see Fig. \ref{fig:2}a) with the contrast of the readout (which increases for larger $\theta$). We empirically determine that a small delay $\theta = 20^\circ$ that is constant over the full imaging region leads to favorable image quality.

\subsection{Anisotropic thermoelectric transport}
In this section, we prove that the solution \boldjr{} to anisotropic Seebeck transport equations (Eq. \ref{eq:1}) is linear in $S_a - S_b$. The steady state condition $\nabla \cdot \boldsymbol{J} = 0$ imposes the second order linear partial differential equation for the electrochemical potential $\Phi$:
\begin{equation}\label{eq:3}
	\sigma_a \partial_x^2 \Phi + \sigma_b \partial_y^2 \Phi = - (S_a \sigma_a \partial_x^2 T + S_b \sigma_b \partial_y^2 T),
\end{equation}
with boundary equation
\begin{equation}\label{eq:4}
\hat{n} \cdot \boldsymbol{J} = 0
\end{equation}
on the sample edges. Since Eq. \ref{eq:3} is linear, we separate the source term (the right hand side of Eq. \ref{eq:3}) into two parts and look for solutions of the form $\Phi = \Phi_A + \Phi_B$, where
\begin{equation}\label{eq:5}
	\sigma_a \partial_x^2 \Phi_A + \sigma_b \partial_y^2 \Phi_A = - S_b (\sigma_a \partial_x^2 T + \sigma_b \partial_y^2 T)
\end{equation}
and
\begin{equation}\label{eq:6}
	\sigma_a \partial_x^2 \Phi_B + \sigma_b \partial_y^2 \Phi_B = - (S_a - S_b) \sigma_a \partial_x^2 T.
\end{equation}
We choose the current corresponding to the two potentials $\Phi_{A,B}$ to be
\begin{equation}
\begin{split}\label{eq:7}
	J_A,x &= -\sigma_{a}\big(\partial_{x} \Phi_A + S_{a}\partial_{x} T\big)\\
	J_A,y &= -\sigma_{b}\big(\partial_{y} \Phi_A + S_{b}\partial_{y} T\big),
\end{split}
\end{equation}
and
\begin{equation}
\begin{split}\label{eq:8}
	J_B,x &= -\sigma_{a}\partial_{x} \Phi_B \\
	J_B,y &= -\sigma_{b}\partial_{y} \Phi_B,
\end{split}
\end{equation}
such that the total current $\boldsymbol{J} = \boldsymbol{J_A} + \boldsymbol{J_B}$ is unchanged from the original problem (Eq. \ref{eq:1}).

Looking only at Eq. \ref{eq:5} (isotropic part), we have the simple solution $\Phi_A = -S_b T$. This yields the current:
\begin{equation}
\begin{split}\label{eq:9}
	J_A,x &= -\sigma_{a} (S_{a}-S_{b}) \partial_{x} T\\
	J_A,y &= 0.
\end{split}
\end{equation}
Now, the remaining problem is to solve for $\Phi_B$ under Eq.~\ref{eq:6} (anisotropic part), subject to the boundary equation on the total current $\boldsymbol{J}$ (Eq. \ref{eq:4}), which implies:
\begin{equation}\label{eq:10}
\hat{n} \cdot \boldsymbol{J_B} = -\hat{n} \cdot \boldsymbol{J_A}.
\end{equation}
Since $\boldsymbol{J_A} \propto S_a - S_b$ (Eq. \ref{eq:9}), it is clear that both Eqs.~\ref{eq:6} and \ref{eq:10} are linear in $S_{a}-S_{b}$. Hence, the solution $\Phi_B$ and accordingly, $\boldsymbol{J_B}$ must also be proportional to $S_{a}-S_{b}$. Thus, the total current $\boldsymbol{J} = \boldsymbol{J_A}+\boldsymbol{J_B} \propto S_{a}-S_{b}$.

\subsection{Ab initio calculations of anisotropic in-plane thermopower}
Ab initio calculations are performed within the framework of density functional theory, including spin-orbit coupling as implemented in VASP \cite{Kresse1993,Kresse1994a}. We use the Perdew-Burke-Ernzerhof exchange-correlation and the projector augmented wave method with an energy cutoff of 312 eV. The ground state is determined on a 12×10×6 $\Gamma$-centered $\boldsymbol{k}$-grid with a Gaussian smearing of 0.05 eV. A subsequent non-self-consistent calculation is employed to determine the band energies on a fine $\boldsymbol{k}$-grid of 144×120×72, which is used in the transport calculation.

The Seebeck coefficients of \WTe{} along the in-plane $a$- and $b$-axes, $S_a$ and $S_b$, are calculated within the constant relaxation time approximation for the carrier lifetimes, which has previously been found to accurately describe the thermopower of a number of materials \cite{Scheidemantel2003,Madsen2003,Parker2013}. Then, the temperature-dependent thermopower can be expressed as:
\begin{equation}\label{eq:11}
S_{i i}(T)=-\frac{1}{e T} \frac{\int \sigma_{i i}(\varepsilon)(\varepsilon-\mu)\left(-\frac{\partial f^{0}}{\partial \varepsilon}\right) d \varepsilon}{\int \sigma_{i i}(\varepsilon)\left(-\frac{\partial f^{0}}{\partial \varepsilon}\right) d \varepsilon}
\end{equation}
where the transport distribution function is
\begin{equation}\label{eq:12}
 \sigma_{i i}(\varepsilon)=\sum_{n} \int d \mathbf{k}\; v_{n i}^{2}(\mathbf{k}) \delta\left(\varepsilon-\varepsilon_{n}(\mathbf{k})\right).
\end{equation}
Here, $T$ is the temperature, $\varepsilon$ is the carrier energy, $f^0$ is the Fermi distribution function, $\varepsilon_n(\boldsymbol{k})$ gives the energy in band $n$ with wave vector $\boldsymbol{k}$, for which $v_{ni}(\boldsymbol{k})$ is the $i^{th}$ component of the velocity ($i = a$ or $b$). All energies are measured from the temperature-dependent chemical potential $\mu (T)$, and the $\delta$-function in Eq. \ref{eq:12} is handled with the tetrahedron method.  

Calculations are performed using the experimental lattice parameters: a = 3.477~\AA, b = 6.249~\AA, and c = 14.018~\AA\; \cite{Mar1992}. The band structure, Fermi surface, and density of states near the Fermi level of \WTe{} are shown in Extended Data Fig. \ref{ext:3}a-c. The band structure is nearly identical to that obtained in prior first-principles calculations \cite{Soluyanov2015}. The calculated $S_a$ and $S_b$ are plotted as a function of the temperature $T$ in Extended Data Fig. \ref{ext:3}d. Around room temperature, both $S_a$ and $S_b$ are negative, reflecting larger contributions from electrons in the low-lying conduction bands compared with the corresponding contributions from holes occupying the uppermost valence bands. A temperature-dependent anisotropy is evident; for example, $|S_a|$ is approximately five times larger than $|S_b|$ at 300 K. For $T$ increasing above 300 K, $S_a$ becomes more negative while $S_b$ changes sign and becomes increasingly positive. In Supplementary Section 5, we break down in detail the balance between electron and hole contributions to the thermopower that underlie its temperature-dependent behavior.

\subsection{Data availability}
The data that support the findings of this study are available from the corresponding author upon request.

\subsection{Author contributions}
Y.-X.W. and B.B.Z. devised the experiments. Y.-X.W. fabricated the photocurrent devices and performed the SPCM and PCFM experiments. Y.R. conceived and demonstrated the theoretical model, with final implementation by Y.-X.W. C.L. and D.B. performed the ab initio calculations. X.Y. and F.T. synthesized the \WTe{} samples. R.D. and Z.L. synthesized the \TaIrTe{} samples. Y.-X.W. and B.B.Z analyzed the data. Y.-X.W. built the experimental setup, with assistance from X.-Y.Z. and T.K.M.G. B.B.Z., Y-X.W., D.B., and C.L. wrote the manuscript with input from all authors. B.B.Z. supervised the project.

\subsection{Competing interests}
The authors declare no competing interests.

\clearpage

\renewcommand{\figurename}{Extended Data Fig.}
\renewcommand{\thefigure}{\arabic{figure}}
\renewcommand{\theHfigure}{Extended Data Fig. \arabic{figure}} 
\setcounter{figure}{0}

\begin{figure*}[t]
\includegraphics[scale=1]{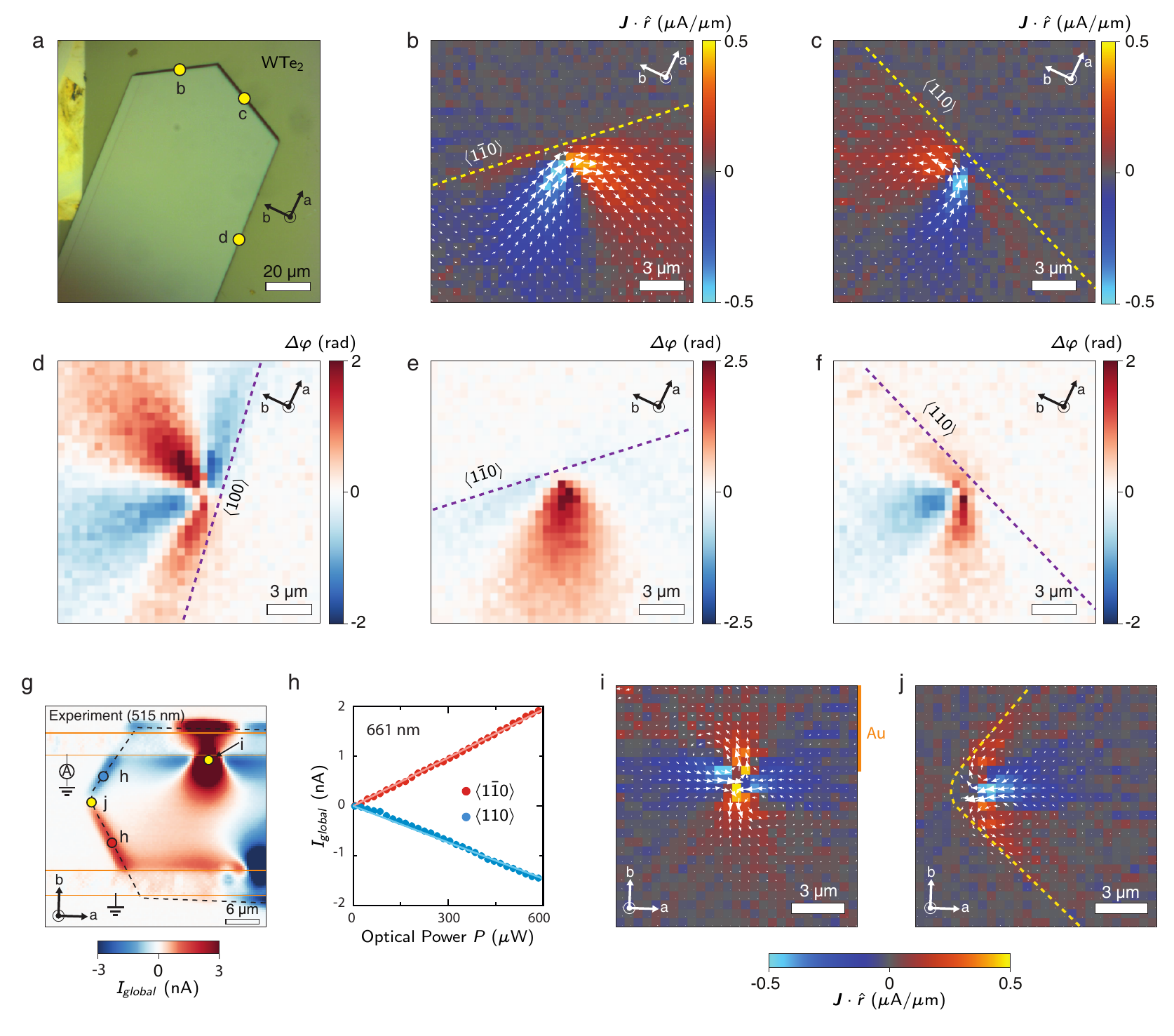}
\caption{\label{ext:1}The APTE and global photocurrent in \WTe{} devices. (a) Zoomed-in optical micrograph of \WTe{} Device A. The photoexcitation locations for the PCFM images to follow are labeled by their figure panel. (b) PCFM image of $\boldsymbol{J}(\boldsymbol{r})$ for photoexciting the $\langle 1\bar{1}0 \rangle$ edge with $P$ = 70 $\mu$W. (c) PCFM image for photoexciting the $\langle 110 \rangle$ edge with $P$ = 120 uW. (d) NV center phase map $\Delta \varphi(\boldsymbol{r})$ for photoexciting the $\langle 100  \rangle$ edge in Device A, corresponding to the reconstructed \boldjr{} shown in Fig. \ref{fig:4}d of the main text. (e) Phase map $\Delta \varphi(\boldsymbol{r})$ corresponding to the \boldjr{} image in b). (f) Phase map $\Delta \varphi(\boldsymbol{r})$ corresponding to the \boldjr{} image in c). (g) Experimental SPCM image of $I_{global}$ in \WTe{} Device B for $P$ = 300 $\mu$W at 515 nm. The collected photocurrents have an identical pattern as the measurement at 661 nm (Fig. \ref{fig:4}a), but have slightly higher intensity likely due to better objective transmission and sample absorption at shorter wavelengths. (h) Linear dependence of $I_{global}$ on the optical power $P$ at 661~nm for fixed photoexcitation locations on the $\langle 1\bar{1}0 \rangle$ (red) or $\langle 110 \rangle$ (blue) edges in Device B. The global electrical measurement is consistent with the local magnetic measurement ($\Delta \varphi$) presented in Fig. \ref{fig:2}b. (i) PCFM image of $\boldsymbol{J}(\boldsymbol{r})$ when photoexciting the region of high interior $I_{global}$ near the upper gold (Au) electrical pad. The four-fold APTE pattern is negligibly changed from the center of the flake. (j) PCFM image for photoexcitation at the corner between the $\langle 110 \rangle$ and $\langle 1\bar{1}0 \rangle$ edges. The locations for the measurements (h-j) within Device B are labeled in the SCPM image shown in g).}
\end{figure*}

\begin{figure*}[t]
\includegraphics[scale=1]{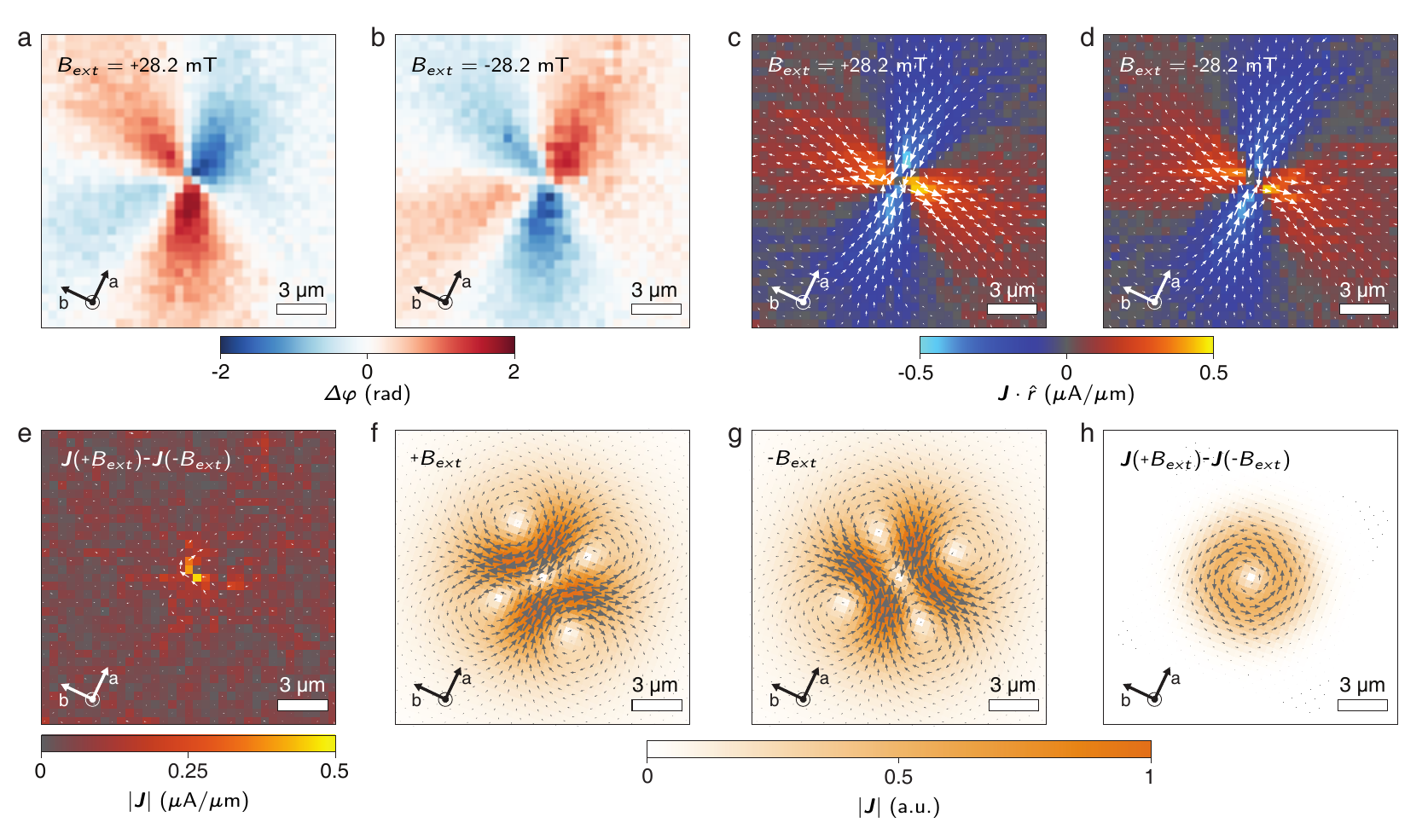}
\caption{\label{ext:2}Absence of any detectable photo-Nernst effect in \WTe{} at room temperature. (a) NV center phase image $\Delta \varphi(\boldsymbol{r})$ for an external dc magnetic field, $B_{ext}$ = +28.2 mT. (b) $\Delta \varphi(\boldsymbol{r})$ image for the opposite field direction, $B_{ext}$ = -28.2 mT. By flipping the direction of $B_{ext}$, we probe the projection of the photocurrent's magnetic field, $B_{PC}$, along opposite NV axes. Afterwards, the NV center's acquired phase simply changes sign everywhere, indicating that the current flow is independent of the magnetic field direction. (c) Reconstructed \boldjr{} for $B_{ext}$ = +28.2 mT. (d) Reconstructed \boldjr{} for $B_{ext}$ = -28.2 mT. (e) The difference between the experimental photocurrent patterns at positive and negative external field: $\boldsymbol{J}(\boldsymbol{r}, +B_{ext}) - \boldsymbol{J}(\boldsymbol{r}, -B_{ext})$. The difference highlights the field-antisymmetric component of the photocurrent flow, which isolates the photo-Nernst effect \cite{Cao2016,Zhou2019}. The overlaid false colormap denotes the magnitude of the difference, which is experimentally consistent with noise. (f,g) Simulated photocurrent pattern $\boldsymbol{J}(\boldsymbol{r})$ when including both the APTE term and a possible Nernst term ($\propto B_{ext,z}\boldsymbol{\hat{z}}\times \nabla T$) for f) positive and g) negative $B_{ext}$ (see Supplementary Section 6). (h) The difference $\boldsymbol{J}(\boldsymbol{r}, +B_{ext}) - \boldsymbol{J}(\boldsymbol{r}, -B_{ext})$ between the simulated patterns clearly shows a chiral photocurrent vortex due to the photo-Nernst effect, which is absent in the experiment. The colormaps in (f-h) denote the magnitude $|\boldsymbol{J}|$.}
\end{figure*}

\begin{figure*}[t]
\includegraphics[scale=1]{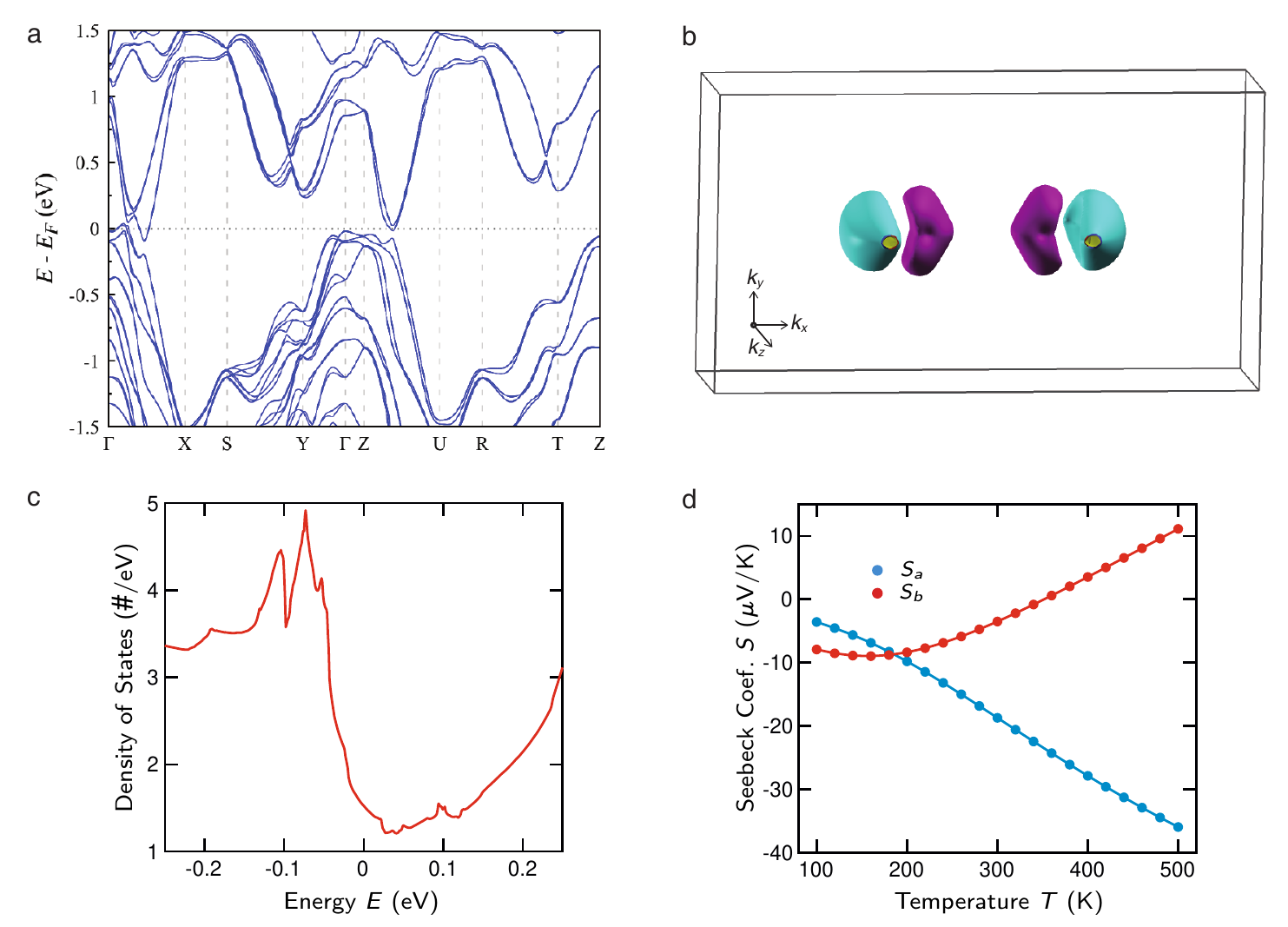}
\caption{\label{ext:3} Ab initio calculations of the band structure and in-plane thermopower of \WTe{}. (a) Electronic band structure calculated within density functional theory, including spin-orbit coupling. (b) Fermi surface of \WTe{} with an outer electron pocket shown in cyan and an outer hole pocket shown in purple. (c) Electronic density of states for \WTe{}, displaying an electron-hole asymmetry about $E = 0$, defined at the chemical potential $\mu$(300 K). (d) Calculated Seebeck coefficients $S_a$ and $S_b$ along the $a$- and $b$-axes of \WTe{} versus temperature $T$. Both $S_a$ and $S_b$ are negative at 300 K, with $S_a < S_b$ as observed in experiment. The difference $|S_a - S_b|$ diverges with increasing temperature up to 500 K: the thermopower along the $a$-axis is dominated by electron contributions due to the high velocities of the low-lying conduction bands along this direction, but the smaller velocities along the $b$-axis do not compensate for the increasing hole contribution that is probed at higher temperatures, due to the rapidly rising density of states on the hole side away from the chemical potential.}
\end{figure*}

\begin{figure*}[t]
\includegraphics[scale=1]{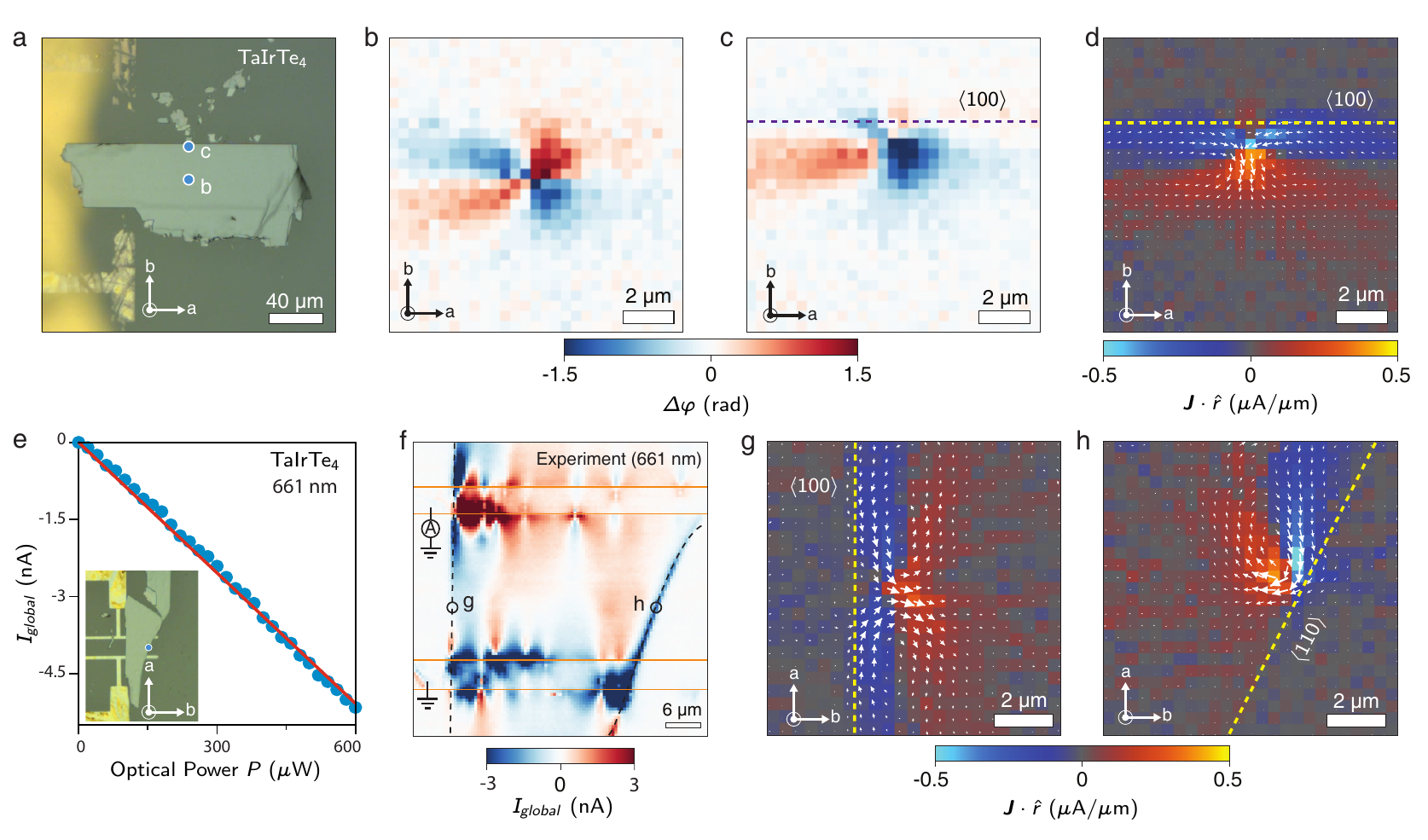}
\caption{\label{ext:4}The APTE and global photocurrent in \TaIrTe{} devices. (a) Optical micrograph of \TaIrTe{} Device A, with thickness 440 nm. (b) NV center phase image $\Delta \varphi(\boldsymbol{r})$ for photoexcitation in the interior of \TaIrTe{} with $P = 90~\mu$W. The reconstructed \boldjr{} from $\Delta \varphi(\boldsymbol{r})$ is shown in Fig. \ref{fig:3}d of the main text. (c) Phase image $\Delta \varphi(\boldsymbol{r})$ for photoexcitation of the $\langle 100 \rangle$ edge in \TaIrTe{} with $P = 90~\mu$W. The photoexcitation locations of b) and c) are labeled in the optical micrograph shown in a). (d) Reconstructed PCFM image of \boldjr{} near the $\langle 100 \rangle$ edge, corresponding to $\Delta \varphi(\boldsymbol{r})$ in c). (e) Dependence of the Shockley-Ramo photocurrent $I_{global}$ on the optical power $P$ at 661 nm for an oblique $\langle 110 \rangle$ edge in \TaIrTe{}. The data are taken on \TaIrTe{} Device B with thickness of 280 nm, whose optical micrograph is shown as the inset. (f) Experimental SPCM image of \TaIrTe{} Device B. Similar to \WTe, global edge photocurrent is detected along the right $\langle 110 \rangle$ edge, while interior photocurrents are detected throughout the device. Far away from the contacts, the edge photocurrent is nearly vanishing along the left $\langle 100 \rangle$ edge, as expected from mirror symmetry and a gradient field $\nabla \Psi$ that becomes slowly-varying over the local photocurrent extent. (g,h) PCFM image of $\boldsymbol{J}(\boldsymbol{r})$ for photoexciting: g) the $\langle 100 \rangle$ edge, and h) the $\langle 110 \rangle$ edge in \TaIrTe{} Device B, both with $P = 85~\mu$W. The photoexcitation locations for g) and h) are labeled on the SPCM image in f).}
\end{figure*}

\begin{figure*}[t]
\includegraphics[scale=1]{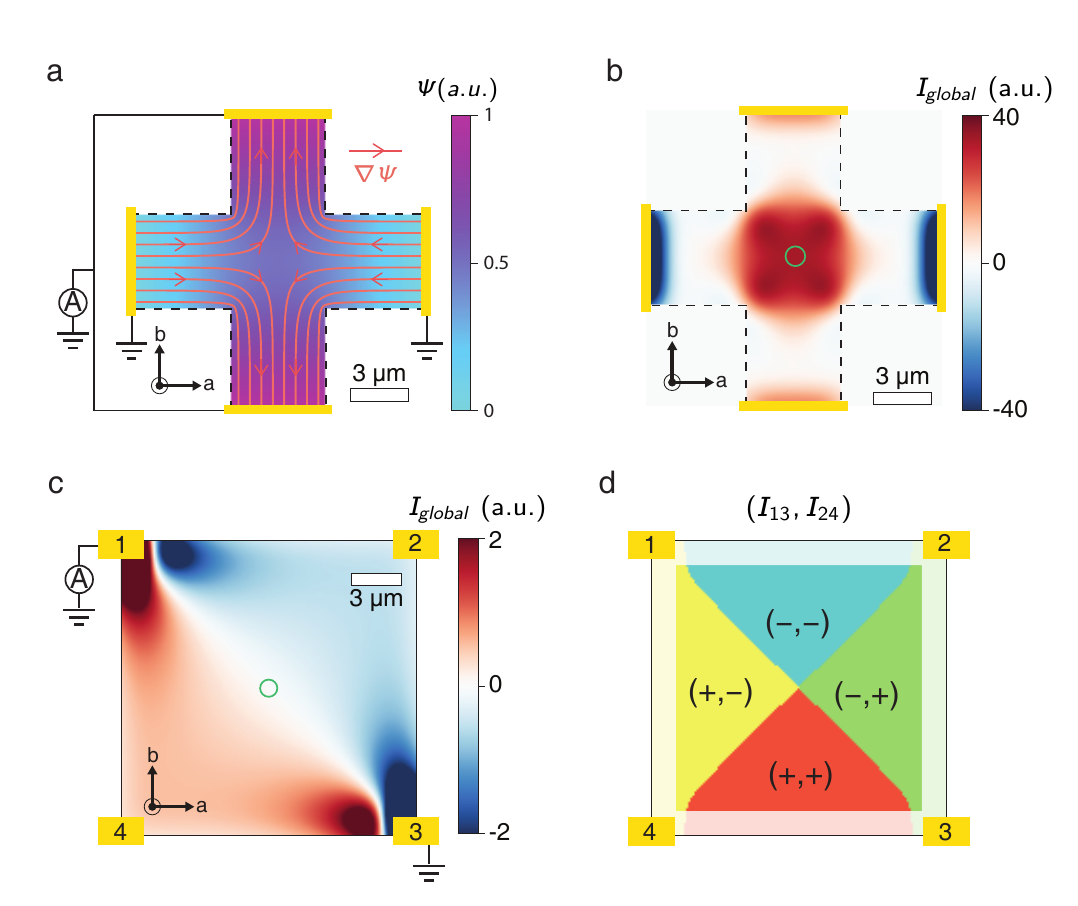}
\caption{\label{ext:5}Novel photodetector designs based on the APTE. (a) Geometry for enhanced Shockley-Ramo collection of bulk APTE photocurrents. We choose the contact configuration to maximize the dot product between $\nabla \Psi$, shown as the streamlines, and $\boldsymbol{J}_{ph}$ due to the APTE in the center of the device. An array of such individual pixels could potentially be used for imaging applications in the mid-infrared or terahertz wavelengths. (b) Simulated $I_{global}$ as a function of the incident beam position for the design in a). The beam size is shown as the green circle and can be realized with an attached microlens. (c) Single-chip, four-quadrant APTE photodetector based on sign-switching, crystal-axes-aligned APTE photocurrents. Photocurrents $I_{13}$, between contacts 1 and 3, or $I_{24}$, between contacts 2 and 4, are alternatively measured with the other contact pair floating. The modeled beam size is shown as the green circle. The material and beam properties are the same between designs shown in b) and c), which allows a comparison of the amplitude of $I_{global}$. (d) The binary pair corresponding to the signs of $I_{13}$ and $I_{24}$ uniquely identifies on which quadrant of the detector the beam is incident. The orientation of the crystal $a/b$-axes for each design is diagrammed.}
\end{figure*}

\end{document}